# GALAXY FORMATION FROM SPECTROSCOPY OF EXTRAGALACTIC GLOBULAR CLUSTER SYSTEMS [1]

Stephen E. Zepf
*Department of Astronomy, University of California*
*Berkeley, CA 94720, USA*

## ABSTRACT

Abstract. I discuss how spectroscopy of extragalactic globular clusters provides a powerful probe of the formation history and mass distribution of galaxies. One critical area is spectroscopy of objects which have been identified as candidate young globular clusters through HST imaging of galaxy mergers. I discuss how such data can constrain models of globular cluster and galaxy formation. As an example, I present new spectra which confirm the presence of young globular clusters in NGC 1275. A second way wide-field spectroscopy can be used to probe the formation history and mass distribution of galaxies is through spectroscopy of large numbers of globular clusters around elliptical galaxies. Metallicities obtained from such data place strong constraints on models of galaxy formation, and velocities determined from the same data provide kinematical tracers of the mass distribution out to distances of $\sim$ 100 kpc.

## 1. Introduction

Globular clusters have long been used as important tracers of the history of chemical enrichment and mass distribution in our Galaxy and in the Local Group. Technological advances have now enabled these techniques to be applied to more distant galaxies. This paper concentrates on two ways spectroscopic study of extragalactic globular clusters has recently begun to constrain the formation history and mass distribution of elliptical galaxies. The first of these is spectroscopy of candidate young globular clusters discovered in high resolution images of interacting and merging galaxies. The second is spectroscopy of the globular cluster systems around elliptical galaxies. I discuss how the initial results of each of these approaches are consistent with the hypothesis that mergers play an important role in the formation and evolution of elliptical galaxies.

## 2. Globular Cluster Formation in Galaxy Mergers

Some environments are much more favorable for the formation of globular clusters than others. For example, the disks of undisturbed spiral galaxies appear to be inhospitable to globular cluster formation, as evidenced by the absence of young globulars in the thin disks of the Galaxy and M31. More generally, spiral galaxies have fewer globular clusters per unit luminosity or mass than elliptical galaxies (Harris 1991; Zepf & Ashman 1993). This result suggests that the conditions during

---

the formation of elliptical galaxies were more conducive to the formation of globular clusters than those typical of the disks of spiral galaxies.

Ashman & Zepf (1992) proposed that the richer globular cluster populations of ellipticals relative to spirals can be understood if elliptical galaxies form from the mergers of gas-rich disk galaxies, and that globular clusters are formed during these mergers. They and many others (e.g. Schweizer 1987, Larson 1990, Kumai et al. 1993, Harris & Pudritz 1994) have argued that the physical conditions expected in such mergers are favorable for globular cluster formation. Ashman & Zepf (1992) therefore predicted that if elliptical galaxies form by merging, newly formed globular clusters should be observable in tidally interacting and merging galaxies.

This prediction received dramatic support from HST imaging of the peculiar galaxy NGC 1275, which revealed objects with the luminosity, color, and size expected of young globular clusters. The success of the Ashman-Zepf prediction was demonstrated even more strongly in the observation of similar bright, blue compact sources in the prototypical merging system NGC 7252 (Whitmore et al. 1993). These observations of roughly 100 young clusters in several galaxy mergers provide strong evidence that the physical conditions in tidally interacting and merging systems are favorable for globular cluster formation. This conclusion had previously been hinted at by the existence of at least a few young globular clusters in the LMC (e.g. Mateo 1993) and by the super-star clusters in several dwarf irregulars, which often appear to be interacting (e.g. Kennicutt & Chu 1988). In this context, a young globular cluster is an object that after $\sim 10$ Gyr of stellar evolution will have the properties characteristic of Galactic globular clusters.

Although HST imaging provides a strong argument that the bright, blue, compact objects observed in NGC 1275 and NGC 7252 are young globular clusters, spectroscopy of these objects is the final, critical step in confirming such an identification. The most basic aim of this spectroscopy is to confirm for individual objects that they are associated with the merging galaxy. The second critical component is to confirm that the optical emission of these objects comes from stars and therefore that the models which transform a young cluster's luminosity and color to mass are at least roughly valid. With good spectra, it is also possible to use the strength of various absorption lines to provide better constraints on the ages of the young globulars. Better ages lead directly to improved estimations of the mass from stellar population models. Finally, with multi-object techniques, large telescopes, and excellent image quality, it will be possible to obtain spectra and determine velocities for a number of clusters. This will allow at least a rough determination of the kinematics of the young cluster population.

In order to obtain spectroscopy of the candidate young globular clusters discovered in NGC 1275, we used the LDSS-2 on the WHT in October of 1993. Because of the excellent seeing and good perfomance of the spectrograph, the most luminous candidate clusters were clearly identifiable above the bright galaxy background. For the brightest of the candidate clusters, we have been able to obtain good spectra at two position angles. In Figure 1, we show one of these spectra, and for comparison, a spectrum of an A star obtained during the same night. This figure clearly demon-strates that the bright, blue object seen in the HST images is in fact a young star cluster in the galaxy NGC 1275. A more detailed analysis of this data is presented in Zepf et al. (1994).

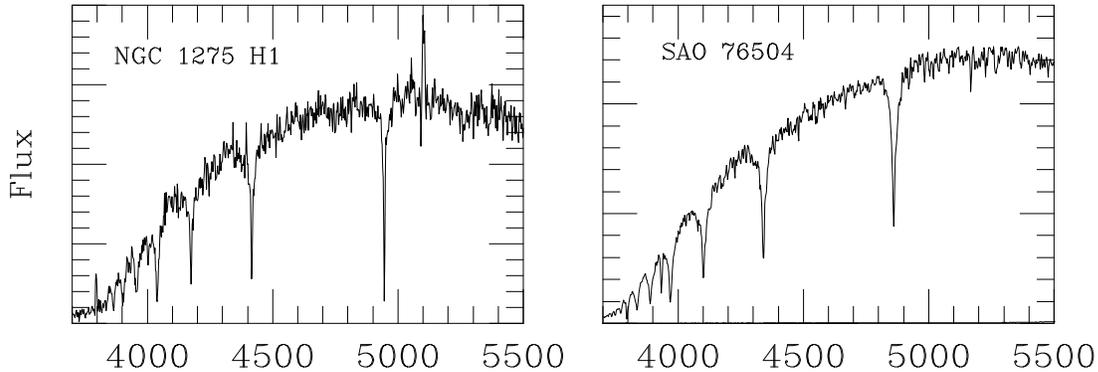

Figure 1 - The spectrum on the left is of the brightest object in the Holtzman et al. (1992) list of candidate young clusters, and the spectrum on the right is of the A star SAO 76504.

This spectral confirmation of the existence of young globular clusters in NGC 1275 is further evidence that globular clusters can form in tidal interactions and mergers. These data suggest that galaxy mergers are fertile ground for studying the astrophysics of globular cluster formation. Moreover, this result is consistent with the hypothesis that the greater specific frequency of globular clusters around ellipticals relative to spirals is the result of the formation of globular clusters in the mergers which make the elliptical galaxies. The next step is to test this hypothesis in more detail by determining the efficiency with which globular clusters form in mergers of various types.

## 3. Globular Cluster Systems of Elliptical Galaxies

Globular cluster systems are invaluable probes of the formation history and mass distribution of galaxies. For all but a few nearby galaxies, globular clusters provide the most observationally accessible way to study the ages, metallicities and kinematics of individual objects, rather than integrated properties. Since globular clusters are bound, coeval, and chemically homogenous (at least to first order), they provide a distinct record of the physical conditions at the time of their formation.

Because of this property of globular clusters, they can be used to test competing theories of the formation of elliptical galaxies (Ashman & Zepf 1992, Zepf & Ashman 1993). If elliptical galaxies form in a monolithic collapse, the metallicity distribution is generally expected to be smooth and single-peaked. More specific predictions can be made in the context of various models (e.g. Arimoto & Yoshii 1987, Matteuci & Tornambè 1987). In contrast, if elliptical galaxies form through mergers, their globular cluster systems will be a composite of at least two populations. These are the globular clusters associated with the progenitor spirals, and those formed during the merger itself. Since the globulars formed during the merger are formed from enriched disk gas, they will generally be of higher metallicity than the clusters associated with the halos of the progenitor spirals. As a result, the metallicity distribution of the globular systems of ellipticals formed by mergers is expected to have at least two peaks.

The difference between the metallicity distribution of the globular clusters predicted by the monolithic collapse model and by the merger model provides a test of which theory more correctly describes the formation of elliptical galaxies. An observationally efficient way to estimate the metallicity distribution is to study the color distribution of the globular clusters, since broadband colors are primarily driven by metallicity in old stellar systems. In Zepf & Ashman (1993), we first performed this test on the globular cluster systems of the elliptical galaxies NGC 4472 and NGC 5128, the systems with the best photometric data then available in the literature (Couture et al. 1991 for NGC 4472 and Harris et al. 1992 for NGC 5128). Using the KMM algorithim to analyze these distributions (cf. Ashman, Bird, & Zepf 1994), we found they were better fit by a distribution with two peaks than a single one at confidence levels of 98% and 95%, respectively. We and others have gone on to obtain better data for the globular cluster systems of other elliptical galaxies. Six elliptical galaxies have now been analyzed in this way, and all appear to have color distributions which are better fit by models with two or more peaks than single-peaked ones (Zepf, Ashman & Geisler 1994, Secker et al. 1994, Lee & Geisler 1993, Ostrov, Forte, & Geisler 1992).

Although the color distributions provide significant evidence that elliptical galaxies formed through a merging process, the case for or against a merging origin can be made considerably stronger when spectra are obtained for the globular clusters. Firstly, the contamination from foreground stars and background galaxies can be eliminated directly, rather than by the statistical estimates required when only photometry is available. Secondly, the metallicity of the globular clusters can be estimated from absorption-line indices, and compared to the metallicity estimates derived from the broadband colors. Thirdly, comparison of absorption-line indices arising from different elements can provide information about abundance ratios and therefore on the history of chemical enrichment.

The spectroscopic study of large numbers of globular clusters around elliptical galaxies is aided greatly by the close match between the typical angular extent of rich globular cluster systems and the field of view of the latest generation of multislit spectrographs. An example of this good match is shown in Figure 2, which is a plot of the surface density of photometric globular cluster candidates around the elliptical galaxy NGC 3923 (see Zepf, Geisler, & Ashman 1994). This plot demonstrates that for this typical globular cluster system of a bright elliptical at a distance of 1600 $\mathrm{km\,s^{-1}}$, the derived surface density of globular clusters with R $\lesssim$ 22 is a factor of several greater than the estimated background at the last data point at a radius of 5.6 arcminutes.

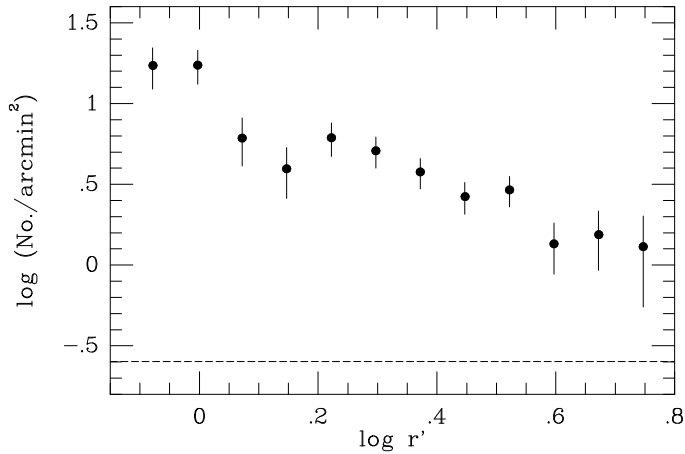

Figure 2 - A plot of the radial profile of the globular cluster system of NGC 3923 for a sample with $20.4 < (T_1)_0 < 22.1$, where $T_1 \sim R$. The estimated background for this photometrically selected sample is given as the dashed line.

This extended spatial distribution, characteristic of the rich globular cluster systems of bright ellipticals, also makes globular cluster excellent kinematic tracers at large galactic radii. This is perhaps the most exciting prospect for wide-field spectroscopy of extragalactic globular cluster systems. Using NGC 3923 as an example, about 100-150 clusters are expected to a limit of $R \sim 22$ within the annular region from 25 $h^{-1}$kpc to 50 $h^{-1}$kpc. The background contamination in a sample selected from images like ours of NGC 3923 is expected to be roughly 50% for these limits.

## 4. Acknowledgements

I thank my collaborators, Keith Ashman, Dave Carter, Doug Geisler, and Ray Sharples for their valuable contributions to the projects described above. I acknowledge support from NASA through grant number HF-1055.01-93A awarded by the Space Telescope Science Institute, which is operated by the Association of Universities for Research in Astronomy, Inc., for NASA under contract NAS5-26555.